\documentclass{article}
\usepackage{graphics}
\usepackage{mark}

\def\ltsima{$\; \buildrel < \over \sim \;$}
\def\simlt{\lower.5ex\hbox{\ltsima}}
\def\gtsima{$\; \buildrel > \over \sim \;$}
\def\simgt{\lower.5ex\hbox{\gtsima}}
\def\etal{{\em et al.}}
\def\hmpc{\,h^{-1} {\rm Mpc}} 
\def\kms {\,{\rm km\,s^{-1}}}
\def\brel{ b_{\rm rel}} 
\def\Sd {S_{\rm d}}
\def\Nc {N_{\rm c}} 
\def\like {{\cal L}}
\def\likemax { {\cal L}_{\rm max} }
\def\btil {\tilde{b}} 

\begin{document}

\title[A Comparison of PSCz and Stromlo-APM Redshift Surveys]
{A Comparison of the PSCz and Stromlo-APM Redshift Surveys}
\author[M.D. Seaborne et al.] 
   {M.D. Seaborne$^1$, W. Sutherland$^1$, 
   H. Tadros$^1$,    G. Efstathiou$^{2}$,
    C.S. Frenk$^{3}$,   \newauthor  
    O. Keeble$^{4}$, S. Maddox$^{2}$, 
    R.G. McMahon$^{2}$, S. Oliver$^{4}$,   
   M. Rowan-Robinson$^{4}$, \newauthor
    W. Saunders$^{5}$,  S.D.M. White$^{6}$ \\
$^1$ Dept. of Physics, Keble Road, Oxford OX1 3RH, UK \\
$^2$ Institute of Astronomy. Madingley Road,     Cambridge CB3 0HA, UK\\
$^3$ Dept. of Physics, South Road, Durham DH1 3LE, UK \\
$^4$ Astrophysics Group, Imperial College,  Blackett Laboratory,
     Prince Consort Road, London SW7 2BZ, UK \\
$^5$ Royal Observatory, Blackford Hill, Edinburgh, EH9 3HJ, UK \\
$^6$ Max Planck Institute for Astrophysik,  Karl-Schwarzschild-Strasse 1,
   D-8046 Garching bei Munchen, Germany.
        }  

\date{WJS, 20 Jan 98.}

\maketitle

\begin{abstract}

We present a direct comparison of the clustering
properties of two redshift surveys covering a common volume of space: 
the recently completed \emph{IRAS}
Point Source Catalogue redshift survey (PSCz) containing 14500
galaxies with a limiting flux of 0.6 Jy at 60 $\mu$m, and the 
optical Stromlo-APM survey containing 1787 galaxies in a region  of
4300 deg$^2$ in the south Galactic cap. 
We use three methods to compare the clustering properties:
the counts-in-cells comparison of Efstathiou (1995, hereafter E95), 
the two-point cross correlation function, and the
Tegmark  (1998) `null-buster' test. 
We find that the Stromlo variances
are systematically higher than those of PSCz, as expected due to
the deficit of early-type galaxies in \emph{IRAS} samples.
However we find that the differences between the cell counts are consistent
with a linear bias between the two surveys, with a relative
bias parameter $\brel \equiv b_{\rm Stromlo}/b_{\rm PSCz} \approx 1.3$
which appears approximately scale-independent.   
The correlation coefficient $R$ between optical and \emph{IRAS} densities
on scales $\sim 20 \hmpc$ is $R \ge 0.72$ at 95\% c.l., 
placing limits on types of `stochastic bias' which affect 
optical and \emph{IRAS} galaxies differently. 
\end{abstract}


\section{Introduction}

Observations of galaxy clustering are a very valuable probe
of cosmology, since they can provide information both on 
the shape of the initial density fluctuations and on the cosmological
parameters such as $\Omega_0$, the composition of the dark
matter etc. 
When combined with related observables such as CMB anisotropy
and galaxy peculiar motions, it should in future be possible
to perform consistency checks since there are 3 observable
functions (the CMB spherical harmonics and the power spectra of
density and velocity fields) predicted by one input function
(the initial power spectrum) 
and a handful of cosmological parameters. 

Following the pioneering studies by Peebles and coworkers in 
the 1970s, the field has undergone rapid growth, with many
large galaxy surveys now in existence, notably the 2-dimensional
APM Galaxy Survey (Maddox \etal\ 1990), the CfA-2 redshift
survey (e.g. Vogeley \etal\ 1992), 
the Las Campanas redshift survey (Shectman \etal\ 1996)
and the \emph{IRAS} PSCz survey (Saunders \etal\ 1997). 

The galaxy surveys obviously measure the distribution of
luminous matter, which probably does not exactly trace the total mass 
distribution which is dominated by dark matter. For instance,
it was shown by Kaiser (1984) that if galaxies form at peaks
of a Gaussian random field, they will be more strongly clustered
than the underlying mass, dubbed biasing. 
Thus, understanding the relationship between galaxies and mass
is important both for estimation of cosmological parameters and
for probing the physics of galaxy formation. 
A common assumption is `linear biasing' given by 
$\delta_g = b \delta_m$ where $\delta_g, \delta_m$ are the 
fractional overdensities relative to the mean  
in galaxies and mass respectively,
and $b$ is a constant `bias parameter'; this assumption
must be inaccurate on small scales for $b > 1$ 
since $\delta_g \ge -1$. However, we may still define
a bias parameter $b(r)$ by e.g. $\xi_{gg}(r) = b(r)^2 \xi_{mm}(r)$, 
and if the biasing is ``local'' in the sense that
the galaxy density is a function only of the mass density smoothed
on small scales, it can be shown that
$b(r)$ must approach a constant on large scales (e.g. Coles 1993, 
Pen 1998).   
A detailed account of the statistics of biasing including
possible non-linearities and stochastic variations, 
and the relationships between the several possible definitions of $b$, is 
given by Dekel \& Lahav (1998). 
Recently, there have been  a number of predictions of bias 
assuming that galaxies correspond to dark matter halos, 
e.g. the analytic model of Mo \& White (1996), which is compared
with N-body simulations by e.g. Jing (1998) and Kravtsov \& Klypin (1998). 
The bias factor assuming various simple prescriptions for
the morphology-density relation is investigated 
by Narayanan \etal\ (1998). 

Surveys of galaxy peculiar motions can directly 
map the mass distribution in the local universe, 
but the error bars per galaxy are large and so the method is limited
to rather local volumes with heavy smoothing. 
Since it is difficult to measure the mass distribution directly, 
another useful probe is to compare the clustering properties
of galaxy surveys with different selection criteria: if 
these cluster differently, at least one class cannot exactly follow
the mass distribution. However, if both classes obey the 
linear bias relation with different values of $b$, 
there is clearly a linear relation between the 
two galaxy density fields and thus the cross correlation 
function $\xi_{12}(r) = \sqrt{\xi_1(r) \xi_2(r)}$ where
$\xi_1, \xi_2$ are the usual autocorrelation functions for 
the two galaxy classes. Thus, if this relation 
does not hold, then the linear-bias model must fail
for at least one of the galaxy classes. 

It is well known (e.g. Dressler 1980;  Guzzo \etal\ 1997) that the fraction of
elliptical galaxies increases with local galaxy density, and
it is also known that \emph{IRAS} galaxies (selected at 60$\mu$m)
have a lower correlation amplitude on small scales than
optically selected galaxies (e.g. Saunders \etal\ 1992; 
Loveday \etal\ 1995). 
It is also known that the density
fields of optical and \emph{IRAS} galaxies are in quite good qualitative
agreement, i.e. the prominent nearby structures such as Virgo, 
Perseus-Pisces, Hydra-Centaurus are common to both types of survey, 
but quantitative comparisons so far have been rather limited
in volume since the 
differing depths and sky coverages cause limited overlap
between the different surveys (e.g. Oliver \etal\ 1996). 
Probably for this reason, there has been considerable scatter
in previous estimates of the relative bias $b_O/b_I$
(e.g. Lahav, Nemiroff \& Piran 1992). 
A comparison of
{\em velocity} fields predicted from \emph{IRAS} 1.2 Jy and
ORS surveys is given by Baker \etal\ (1998), who find very
good agreement between the two velocity fields
for a relative bias factor $\brel = b_O/b_I \approx 1.4$. 
In this paper we provide a comparison of density fields 
over a larger volume of space than previously available, 
using the newly-completed \emph{IRAS} PSCz redshift survey 
and the sparse-sampled Stromlo-APM optical survey. 

The plan of the paper is as follows: 
we summarise  details of the two surveys in \S2, 
and then compare the clustering with three different
statistical methods: a counts-in-cells comparison in \S3,
two-point correlation functions in \S4 and the
`modified $\chi^2$' statistic of Tegmark (1998) in \S5. 
We summarise the conclusions in \S6.

\section{The PSCz and Stromlo-APM redshift surveys}

\begin{figure}
\raisebox{70mm}{\rotatebox{270}{\scalebox{0.32}{\includegraphics[100mm,100mm]%
{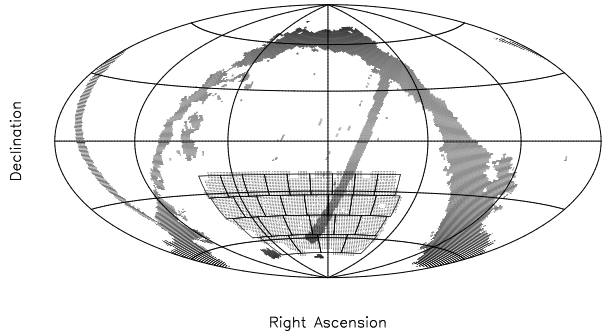}}}}
\caption{An Aitoff projection of the PSCz and Stromlo-APM regions in
equatorial coordinates.  The PSCz mask is indicated by the dark
shading. The light shaded area in the southern hemisphere is the
Stromlo survey region. An example of one of the shells of cells is
also shown. \label{skyplot}}
\end{figure}

The PSCz survey is described in detail by Saunders et al.\ (1997). The
aim of the survey was to obtain redshifts for all \emph{IRAS} galaxies
with 60$\mu$m fluxes greater than 0.6 Jy over as much of the sky as
possible. The parent catalogue is based on the QMW \emph{IRAS} Galaxy
Catalogue (Rowan-Robinson et al.\ 1990) but with a number of
modifications designed to increase sky coverage and improve
completeness. 
The final 2-D catalogue contains 17060 objects, 
1593 of which were rejected as sources in our Galaxy or multiple 
entries from resolved nearby galaxies. 
Another 648 were rejected as faint galaxies or no optical
identification leaving 14819 in the target list. Redshifts are now
known for 14539 (95\%) of these. The PSCz mask is shown in figure
\ref{skyplot} as the dark shaded region. It includes regions of low
galactic latitude, some odd patches of galactic cirrus and the two
strips of ecliptic longitude not surveyed by \emph{IRAS}.

\begin{figure*}

\vspace*{15.0cm}
\includegraphics{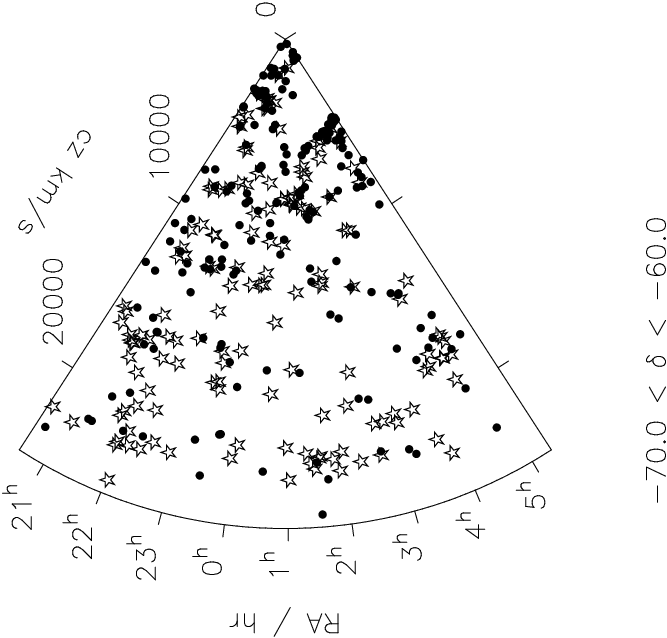}
\includegraphics{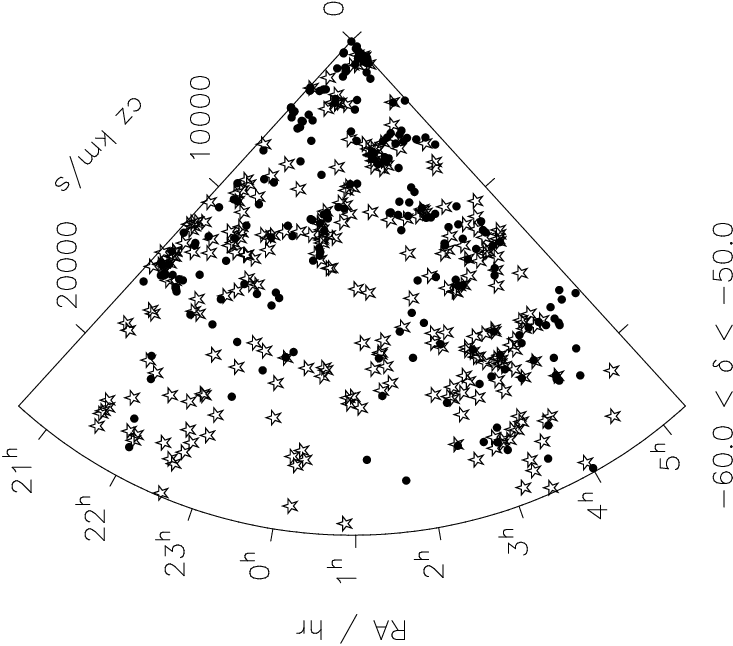}
\includegraphics{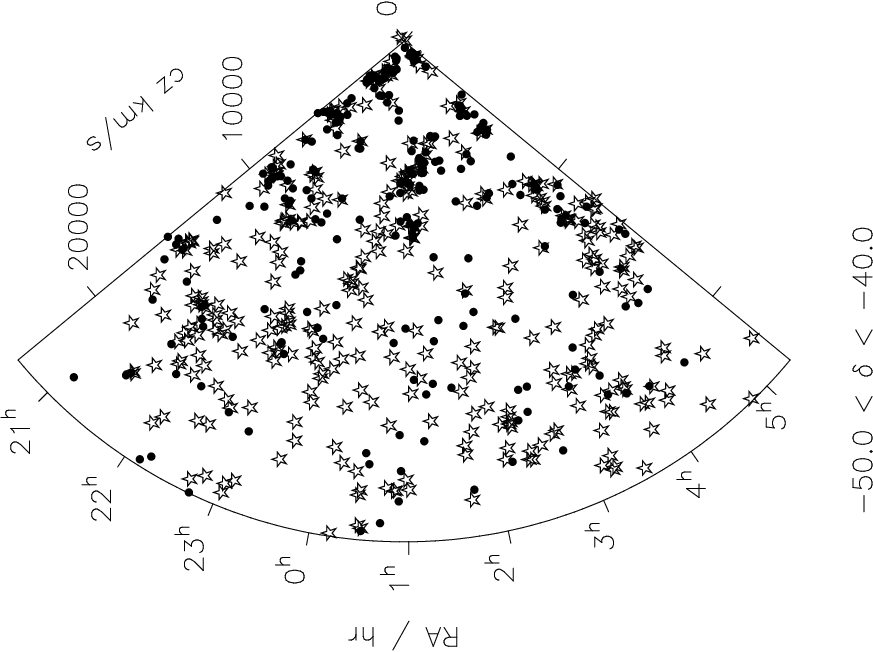}
\includegraphics{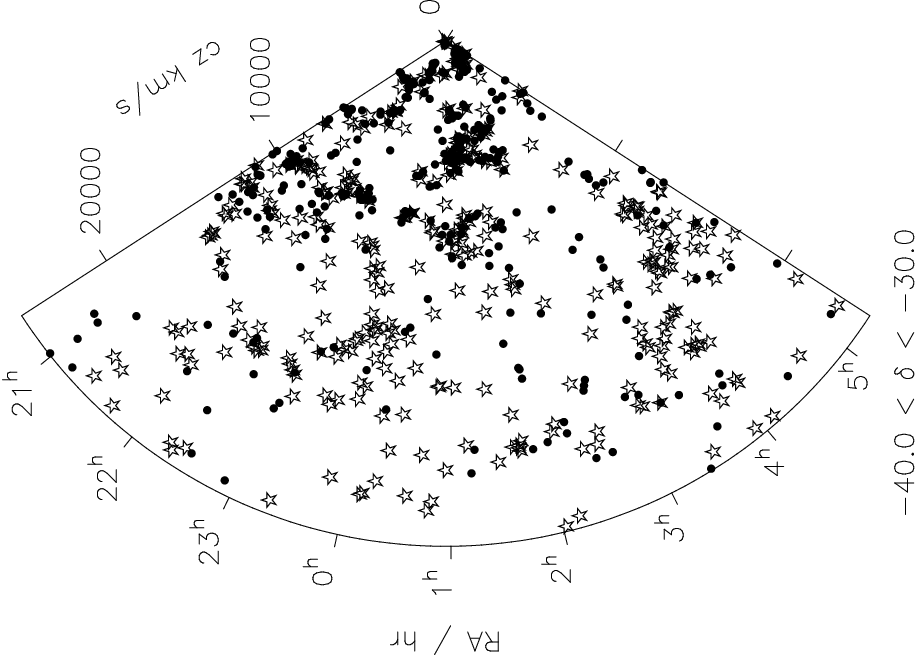}
\vspace*{5mm}

\caption{The locations of galaxies in redshift space at slices of
declination within the APM region. PSCz galaxies are represented by
black spots and Stromlo-APM galaxies by stars.}\label{coneplots}
\end{figure*}

The Stromlo-APM survey (Loveday et al.\ 1992) is an optically selected
redshift survey in a region of the southern hemisphere approximately
$21^{\rm h}\la \alpha \la~5^{\rm h}, -72.5\degr \la \delta \la
-17.5\degr$.  The survey consists of 1787 galaxies of magnitude
$b_j\leq 17.15$ selected at a rate of 1 in 20 from the APM galaxy
catalogue. The Stromlo-APM region is shown by the light shading in
figure \ref{skyplot}. The small holes in the survey are mainly due to
bright foreground stars.  Figure \ref{coneplots} shows `cone plots' of
galaxies in declination slices for the two catalogues in the APM
region.  Note that the selection functions are somewhat different,
with $dN/dz$ peaking at $\sim 70 \hmpc$ for PSCz and $\sim 150 \hmpc$
for Stromlo, and the shot noise is appreciable in both surveys due to
the relatively low space density of \emph{IRAS} galaxies and the
1-in-20 sampling for Stromlo.  However, there is quite good
qualitative similarity between the two distributions, e.g. the
prominent `wall' near $\alpha \sim 21^h$ and the squarish void centred
on $\delta \sim -40^o, cz \sim 7000 \kms$.

In the following analysis we treat the two galaxy surveys as
independent samples of the same region of the universe. This
assumption would not be valid if there were an appreciable number of
galaxies common to both catalogues. We have found 41 such galaxies and
these have not been included in any of the subsequent analysis.

\subsection{Mock PSCz and Stromlo-APM catalogues}
\label{sec-sims} 
Throughout this paper we used a suite of mock PSCz and Stromlo-APM 
surveys to estimate the uncertainty in the various statistics calculated. 
There are 27 pairs of mock catalogues in total taken from  N-body
simulations of standard $\Omega_0=1$ CDM universes. They
have the same survey geometry, selection function and sampling rate as
their real counterparts. Each pair of Stromlo and PSCz catalogues was
sampled from a common region of space in the same initial simulation
so that underlying density fluctuations 
in the PSCz mock catalogue are mirrored in
the corresponding Stromlo catalogue. We would therefore expect to see
the clustering in each catalogue of the pair to differ only due to the
shot noise. This property is useful when calculating
the uncertainty in the cross correlation function and the ratio
between the two individual correlation functions. 
Mass points in the simulations are interpreted as galaxies so
the catalogues do not include bias.

\section{Counts in Cells Comparison}\label{sec-cics}

The counts in cells comparison method is discussed in detail in
E95. The surveyed region of space is divided into spherical shells of
thickness $\ell$ centred on the observer. Each shell is then
subdivided into $\Nc$ approximately 
cubic cells each of volume $V=\ell^{3}$. 
For one catalogue, we
represent the galaxy counts in the $i$th cell by $N_{i}$. The expected
mean cell count is denoted by $\langle N_{i} \rangle = \lambda$. The
variance of the counts in cells in excess of Poisson variance is given
by
\begin{equation}
	S_{1} = {1 \over \Nc - 1} \sum_{i} (N_{i} - \bar{N})^{2} -
	\bar{N}\label{eq:s1}.
\end{equation}
where $\bar{N}$ is the mean cell count.
The expectation value of (\ref{eq:s1}) is
\begin{equation}
	\langle S_{1} \rangle = \lambda^2 \sigma_1^2,
	\label{eq:s1ex}
\end{equation}
where $\sigma_{1}^{2}$ is the variance of the underlying density field
on the scale of $\ell^{3}$. This is equal to a volume integral of the
autocorrelation function, $\xi(r_{12})$, over the cell of size $\ell^3$
\begin{equation}
  \sigma_1^2 =
   {1 \over V^2} \int\!\!\int_{V=\ell^3} \xi(r_{12}) \, dV_1 \, dV_2
   \label{eq:xiint}.
\end{equation}
When the underlying density fluctuations are Gaussian the variance of
$S_{1}$ is given by
\begin{equation}
    \mbox{Var}(S_1) = {1 \over \Nc} [2\lambda^2 (1+\sigma_1^2) +
       4\lambda^3\sigma_1^2 +2\lambda^4\sigma_1^4]. 
\label{eq:vars1}
\end{equation}

The statistics $S_1$ and $S_2$ can be found for the two redshift
surveys (the subscripts now refer to the two individual surveys) 
and used to examine the comparative clustering.
If the two surveys occupy a common region of space then the
comparison can be taken a stage further by computing
the covariance, $S_{12}$, of the cell counts:
\begin{equation}
   S_{12} = {1 \over (\Nc-1)} \sum_i (N_i - \bar{N})
   (M_i - \bar{M})
\label{eq:s12}
\end{equation}
where $N_i$ and $M_i$ now represent the counts from surveys 1 and 2
respectively. The underlying covariance $\sigma_{12}^2$ is defined
equivalently to $\sigma_1^2$ so that
\begin{eqnarray}
	\langle S_{12}\rangle & = & \lambda\mu\sigma_{12}^2 \ ,{\rm where} \\
	\sigma_{12}^2 & = & {1 \over V^2} \int\!\!\int_{V=\ell^3}
      \xi_{12}(r_{12}) \, dV_1 \,dV_2    \nonumber
\label{eq:xi12int}
\end{eqnarray}
where $\mu= \langle M_{i} \rangle$, the expected cell mean
cell count from survey two, and $\xi_{12}(r)$ is  the usual two-point 
cross-correlation function. 

To calculate the uncertainty in the statistics $S_1$, $S_2$ and $S_{12}$ we
must first assume a model (e.g.\ Gaussian) for the clustering. The
need for such a model can be avoided if we consider only the
\emph{differences} between the counts from surveys 1 and 2 in each
cell. The normalised mean square difference between the cell counts is
\begin{equation}
       \Sd = {1 \over (\Nc - 1)}
      \sum_i (\bar{M}N_i - M_i \bar{N})^2 - (\bar{M}+\bar{N})\bar{M}\bar{N}.
\end{equation}
If the galaxy distributions in the two catalogues are Poisson point
processes with identical statistical properties then the statistic $\Sd$
will be solely determined by Poisson statistics and its variance will
be given by
\begin{equation}
    \mbox{Var}(\Sd) \approx\lambda\mu
    {\Nc \over (\Nc-1)^2} [\lambda^3 + \mu^3 
       	+ 4\lambda^2\mu^2 + 2(\lambda^3\mu+\lambda\mu^3)]\label{eq:vars}.
\end{equation}
This value is independent of the underlying nature of the galaxy
density field. The statistic $\Sd$ is related to $S_1$, $S_2$ and
$S_{12}$ by 
\begin{equation}
     \Sd = \bar{M}^2 S_1 + \bar{N}^2 S_2 - 2\bar{M}\bar{N}S_{12}
\end{equation}
and its expectation value is therefore
\begin{equation}
    \langle \Sd \rangle = \lambda^2\mu^2
  (\sigma_1^2 + \sigma_2^2 - 2\sigma_{12}^2)\label{eq:sex}.
\end{equation}
This provides an simple way of determining whether the clustering
properties of the two surveys are consistent while avoiding making any
special assumptions about the nature of the clustering. If the two
catalogues sample the same density field then the three correlation
functions $\xi_1$, $\xi_2$ and $\xi_{12}$ will be identical and we
see from equations (\ref{eq:xiint}) and (\ref{eq:xi12int}) that so too
are $\sigma_1^2$, $\sigma_2^2$ and $ \sigma_{12}^2$. The expectation
value of $\Sd$ given in equation~(\ref{eq:sex}) would therefore be zero
within the limits of the sampling error given by equation~(\ref{eq:vars}).

We also investigate the possibility that the galaxies in the two
catalogues are biased tracers of the same underlying density field. In
particular we look at the linear bias model
$(\delta\rho/\rho)_{\rm gal} =
b(\delta\rho/\rho)_{\rm m}$ where the two overdensity fields
are perfectly correlated. If $b_1$ and $b_2$ are the bias parameters
of the two catalogues with respect to the underlying matter
distribution then the linear bias model implies
\begin{eqnarray}
\nonumber	\sigma_1^2 & = & b_1^2\sigma_{\rm m}^2,\\
\nonumber	\sigma_2^2 & = & b_2^2\sigma_{\rm m}^2,\\
		\sigma_{12}^2 & = & b_1 b_2 \sigma_{\rm m}^2.
\end{eqnarray}
where $\sigma_{\rm m}^2$ is the variance of the underlying
density field.  Hence we can estimate the relative bias
$b_1/b_2$. The above clearly implies 
\begin{equation}
	\sigma_{12}^2 = \sigma_1\sigma_2\label{eq:sig12}
\end{equation}
which can be used to test for consistency with linear bias.

The  above equations for the $S_i$'s apply to a single spherical shell
with constant mean densities $\lambda, \mu$. 
For the full surveys, we now consider a series
of $N_{\rm{shell}}$ spherical shells,
each of thickness $\ell$, centred on the observer, subdivided
into cubical cells as before. 
The statistics $S_1^k$, $S_2^k$ and $\Sd^k$ are now computed 
separately for each shell $k$.

When the number of cells, $\Nc$, is large, the central limit theorem
states that the probability distribution of the $S_1^k$, $S_2^k$ and
$\Sd^k$ will tend to a multivariate Gaussian
\begin{equation}
   p(S_1^k,S_2^k,\Sd^k)dS_1^k\, dS_2^k \,d\Sd^k  =  
   {\exp\left( A \right) dS_1^k\, dS_2^k\, d\Sd^k \over
    (2\pi)^{3/2}(\det\mathbf{V})^{1/2} },  
  \label{eq:multigauss}
\end{equation}
where
\begin{equation}
\nonumber   A = -{1 \over 2} \sum_{i,j}
   (S_i-\langle S_i\rangle) (S_j-\langle S_j \rangle) (V^{-1})_{ij}, 
\end{equation}
and $V_{ij}$ is the covariance matrix $V_{ij} =
\mbox{Cov}(S_i^k,S_j^k) \ (i,j=1,2,d)$. Expressions for the elements of
this matrix can be found in the appendix of E95. We form the
combined likelihood over all shells, 
\begin{equation}
     \like \propto  \prod_{k=1}^{N_{\rm shell}} 
    p(S_1^k,S_2^k,\Sd^k)
 \label{eq:likeli}
\end{equation}
which we maximise with respect to $\sigma_1^2$, $\sigma_2^2$ and
$\sigma_{12}^2$. This method automatically assigns a weight to each
shell. Nearby shells contribute little weight because the number of
cells is small and distant shells contribute little information
because the statistics become dominated by shot noise.

We use lines of constant right ascension and declination as the cell
boundaries to match the geometry of the Stromlo-APM survey region. 
An example
of one of the concentric shells of cells is shown in figure
\ref{skyplot}. Each cell at right ascension $\alpha$ and declination
$\delta$ subtends an angle $\Delta\delta$ in declination and has a
right ascension range $\Delta\alpha$ such that
$\Delta\alpha=\Delta\delta/\cos\delta$. The angle $\Delta\delta$ is
chosen at each radial distance to ensure that the sides of the cell
are of length $\ell$, the shell separation. 

We apply a joint PSCz-Stromlo mask to both surveys so that Stromlo
galaxies within the PSCz mask are not included and vice versa.
A number of the cells are partially or totally covered
by the mask. We correct for this by
replacing equation (\ref{eq:s1}) with the expression 
\begin{eqnarray}
   S_i^2 & = & A / B \ \ {\rm where} \\
  A & = & \left({V \over \Nc } \sum_j{ N_j \over V_j }\right)^2 
   \left\{ \sum_j \left( N_j -V_j {\sum_k N_k \over \sum_k V_k} \right)^2 
  \right.   \nonumber \\
  & & \left. - \left[ 1- { \sum_k V_k^2 \over (\sum_k V_k)^2 }\right] 
    \sum_j N_j \right\} \ , \nonumber \\ 
    B & = & \left({\sum_j N_j \over \sum_j V_j}\right)^2 \times \nonumber \\
  & &  \left[ \sum_k V_k^2  -2 {\sum_k V_k^3 \over \sum_k V_k}
    + {(\sum_k V_k^2)^2 \over (\sum_k V_k)^2 } \right] \nonumber
\label{eq:s2mask}
\end{eqnarray}
(Efstathiou et al.\ (1990)) where the sum extends over all cells in
the $i^{th}$ radial shell and $V_j$ is the
volume of cell $j$ not excluded by the mask.
This expression is only valid when a small
fraction of each cell is excluded by the mask. 
Cells that are greater
than $30\%$ covered by the joint mask were therefore not used in the
analysis. To calculate the statistic $\Sd$ and the mean cell counts
$\bar{N}$ and $\bar{M}$ we artificially increase the galaxy counts in
partially filled cells to account for the fraction 
excluded by the mask.

If we wish to use equation (\ref{eq:multigauss}) to calculate the
likelihood then there must be enough cells in a given radial shell for
the central limit theorem to be applicable. Any shells that do not
contain enough cells for this to be the case must be excluded from the
analysis. We have only included shells with 20 or more filled ($\leq
30\%$  masked) cells. The nearest shells are therefore
rejected. Cells beyond a radial distance of $250 \hmpc$ are not
included in the analysis as they contain very few galaxies and so make
a negligible contribution to the final result. The PSCz survey is only
complete out to this distance for $\left|b\right| \ge 10\degr$. We
therefore mask this region from the PSCz (this has no effect on
the `combined' mask since the Stromlo survey is at $b \simlt -40^o$).

\begin{figure}
\raisebox{65mm}{\rotatebox{270}{\scalebox{0.32}%
{\includegraphics[80mm,80mm]{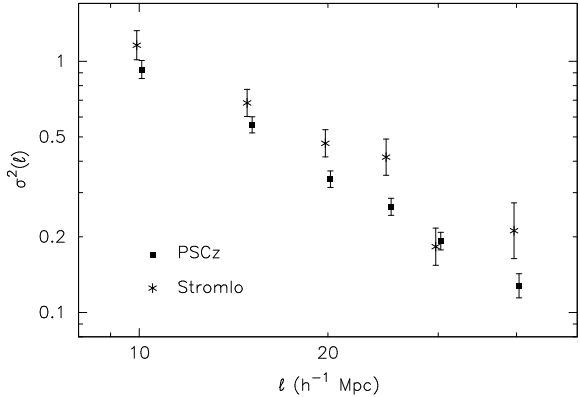}}}}

\caption{$\sigma^2$ as a function of cell size $\ell$ for the PSCz and
Stromlo surveys. The error bars are 1 standard 
deviation. \label{sigsize}}
\end{figure}

Before doing the cell by cell comparison described above we first
measure the values of the cell count 
variances calculated using the whole of
each survey; these are shown as a function of cell size in Figure~
\ref{sigsize}. 
We attempt to minimise the loss of information that arises from binning
the data into cells by shifting the entire grid of cells and
recalculating the likelihood. We shift the grid by 
either 0 or $\ell/2$ from the `default' value
in combinations of right ascension, declination and radial velocity,
thus giving eight separate (although not statistically independent)
estimates of the likelihood. 

For each of the 8 grid positions we form the individual one
dimensional likelihood
\begin{equation}
  \like(\sigma_1^2) = \prod_{k=1}^{N_{\rm shell}}
   \frac{1}{2\pi\mbox{Var}(S_1^i)}
 \exp\left[\frac{-(S_1^i-\bar{N}^2\sigma_1^2)^2}{2\mbox{Var}(S_1^i)}\right]
\end{equation}
We obtain the final estimates of $\sigma_1^2, \sigma_2^2$ 
by computing the product of the 
eight likelihood functions and finding the maximum;
this should provide a better estimate of the peak than using 
a single grid position alone. 
However, it is not
possible to use the combined likelihood to estimate the uncertainty in
the values as the results from each grid position are not independent. 
We generate the error bars in Figure~\ref{sigsize} by repeating the
analysis on the 27 mock PSCz and Stromlo catalogues,
and measuring the standard deviation of the $\sigma^2$ values 
estimated as above.
From the simulations, combining the results from the eight grid positions
gives a 20-40\% gain in precision in the values of
$\log\sigma^2$ over the use of a single set of cells. 
The gain in
precision in general increases with cell size.
The observed sum of the eight $\log \cal{L}$'s is then
re-scaled so that the standard deviation of $\log \sigma^2$ 
matches that from the simulations. 

As expected, the variances decrease with increasing cell size as the
galaxy distribution approaches homogeneity on larger scales, 
and the Stromlo variances are consistently higher than those
of PSCz. For cells of sizes 20, 25 and $40 \hmpc$ the error bars
do not overlap suggesting that the differences in clustering 
amplitude are highly significant. 
Only at $30 \hmpc$ are the values comparable.

\begin{figure}

\hspace{5mm}\raisebox{65mm}{\rotatebox{270}{\scalebox{0.32}[0.29]%
{\includegraphics[110mm,80mm]{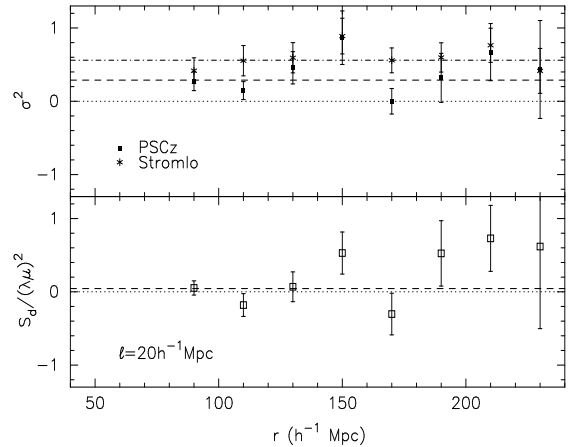}}}}

\caption{A comparison of the counts in $20 \hmpc$ cells for
PSCz and Stromlo galaxies. The top panel shows the values of
$\sigma^2$ for the two catalogues as a function of radial
distance. These are obtained directly from the statistics $S_1$ and
$S_2$. The dashed lines show the respective maximum likelihood values
calculated with the  assumption $\sigma_{12}^2 =
\sigma_1\sigma_2$. The lower panel shows the statistic $\Sd$. The
maximum likelihood value is calculated with the same assumption and
using equation \ref{eq:sex}. All error bars show one standard
deviation. \label{PSCzStromlo20}}
\end{figure}

Figure \ref{PSCzStromlo20} shows the values of $\sigma_1^2$, $\sigma_2^2$
and the statistic $\Sd$ for each radial shell, 
for a cell size of $20 \hmpc$, using
Stromlo and PSCz galaxies within the APM
region only. 

Again we see that the Stromlo variances are, in general, higher. The
maximum likelihood value of the statistic $\Sd$ lies slightly above
zero and has a reduced $\chi^2$ of 1.3 (when compared to the value
zero) suggesting a slight difference in the clustering of galaxies in
the two catalogues.

\subsection{Linear bias} 

\begin{figure}
\hspace{5mm}\raisebox{65mm}{\rotatebox{270}{\scalebox{0.32}[0.29]%
{\includegraphics[110mm,80mm]{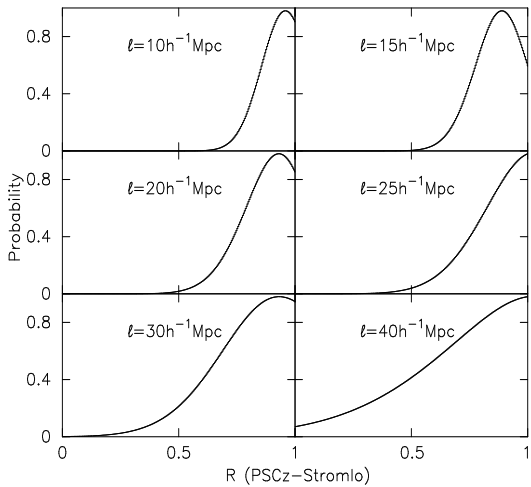}}}}
\caption{The likelihood as a function of the correlation coefficient
$R$.\label{likeliR}}
\end{figure}

We next consider whether these clustering differences are in agreement with
a linear relative bias 
$\sigma_{\rm{optical}} = \brel  \sigma_{\rm{IRAS}}$.  
The likelihood function (equation (\ref{eq:likeli}))
is a function of the three parameters 
$\sigma_1^2$, $\sigma_2^2$ and $\sigma_{12}^2$; 
it is more convenient to change the third parameter to 
the correlation coefficient $R = \sigma_{12}^2/\sigma_1\sigma_2$, 
since we must have $-1 \le R \le 1$ in order that the covariance
matrix in eq.~\ref{eq:multigauss} is positive definite. 
Linear bias clearly implies $R = 1$. 
 We then assume a uniform prior in the 
space of $(\log\sigma_1^2, \log\sigma_2^2, R)$ 
and then integrate ${\cal L}$ over 
the $\log\sigma^2$'s to obtain a one-dimensional
likelihood function for $R$; this is shown in Figure~\ref{likeliR}
for cell sizes in the range $10-40 \hmpc$. 
Again, each value of the likelihood ${\cal L}(R)$ was 
calculated using eight different cell
grid positions, and the values given are the geometric mean of these
eight values. No attempt has been made to reduce the width of the
curves to account for the gain in precision obtained from the use of
multiple grid positions. 

For cells of size $25$ and $40 \hmpc$ the maximum likelihood value of
$R$ occurs at $R = 1$, consistent with linear bias.  Values of $R$
greater than unity are of course unphysical since this would imply
that the correlation between the galaxy counts is better than perfect.
However, it is not unreasonable for the derivative of the likelihood
function to be positive at $R = 1$, if the Poisson differences in the
real cell counts happens to be smaller than the `ensemble average'
value.  For cells of size $10$, $15$, $20$ and $40 \hmpc$ we find that
$R_{\rm{max}}=0.96$, $0.89$, $0.93$ and $0.93$ respectively,
though in each case $R=1$ is well within one standard deviation of this. We
conclude that the galaxy cell counts are consistent with a linear bias
model, and we derive 95\% confidence lower limits of $R \ge 0.83,
0.72, 0.55$ for cells of size $10,20,30 \hmpc$ respectively.

\begin{figure*}

\vspace*{15.3cm}
\includegraphics{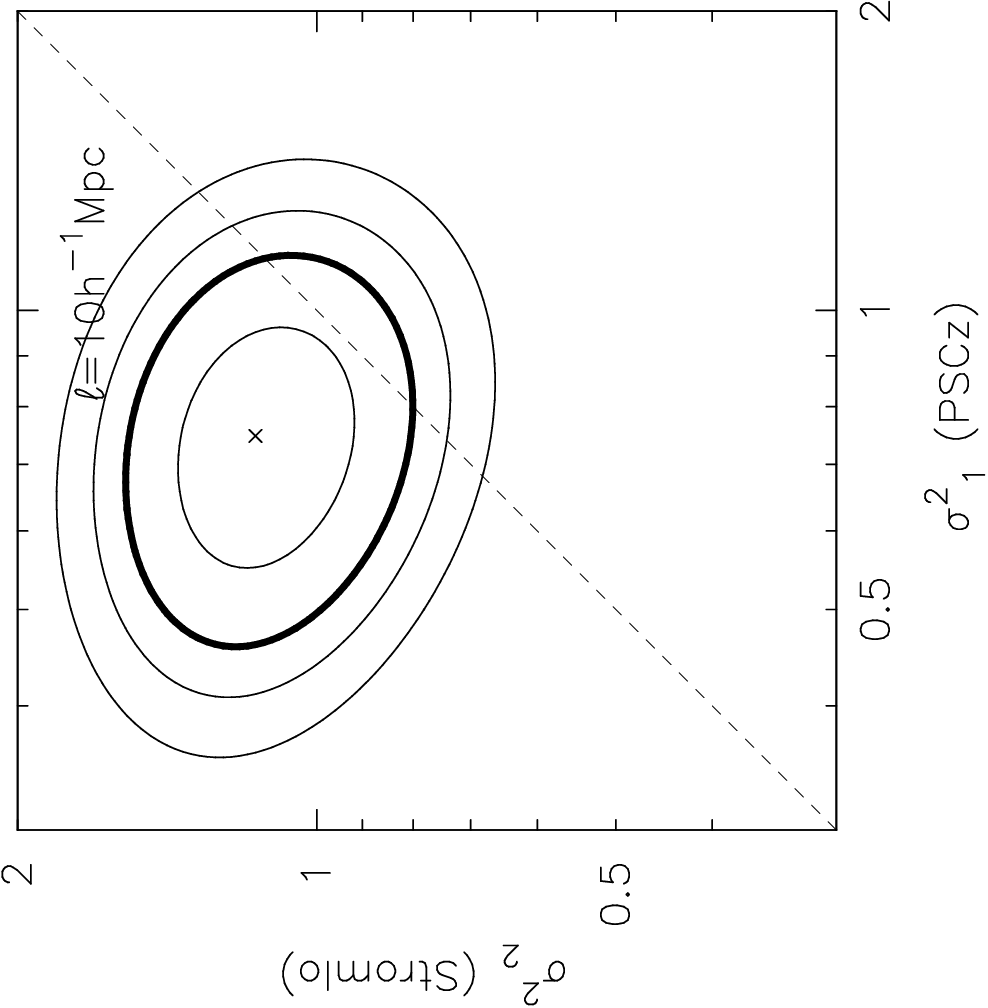}
\includegraphics{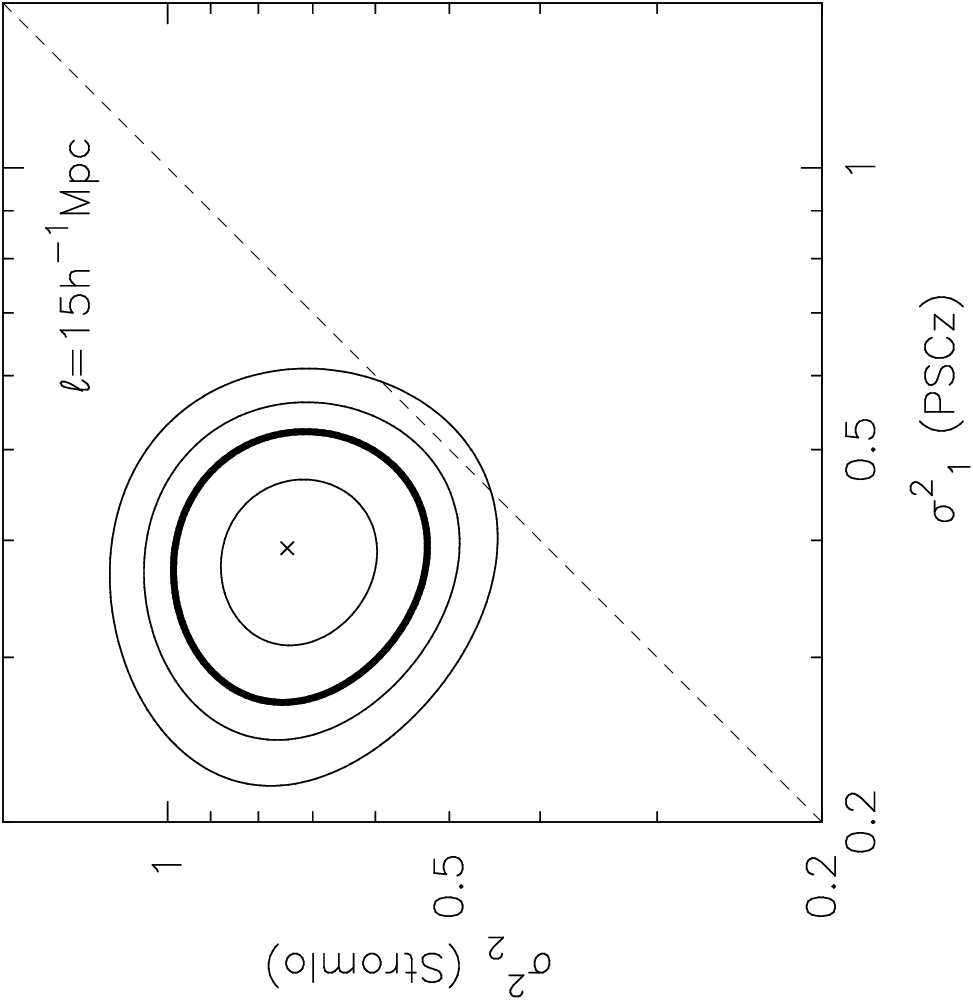}
\includegraphics{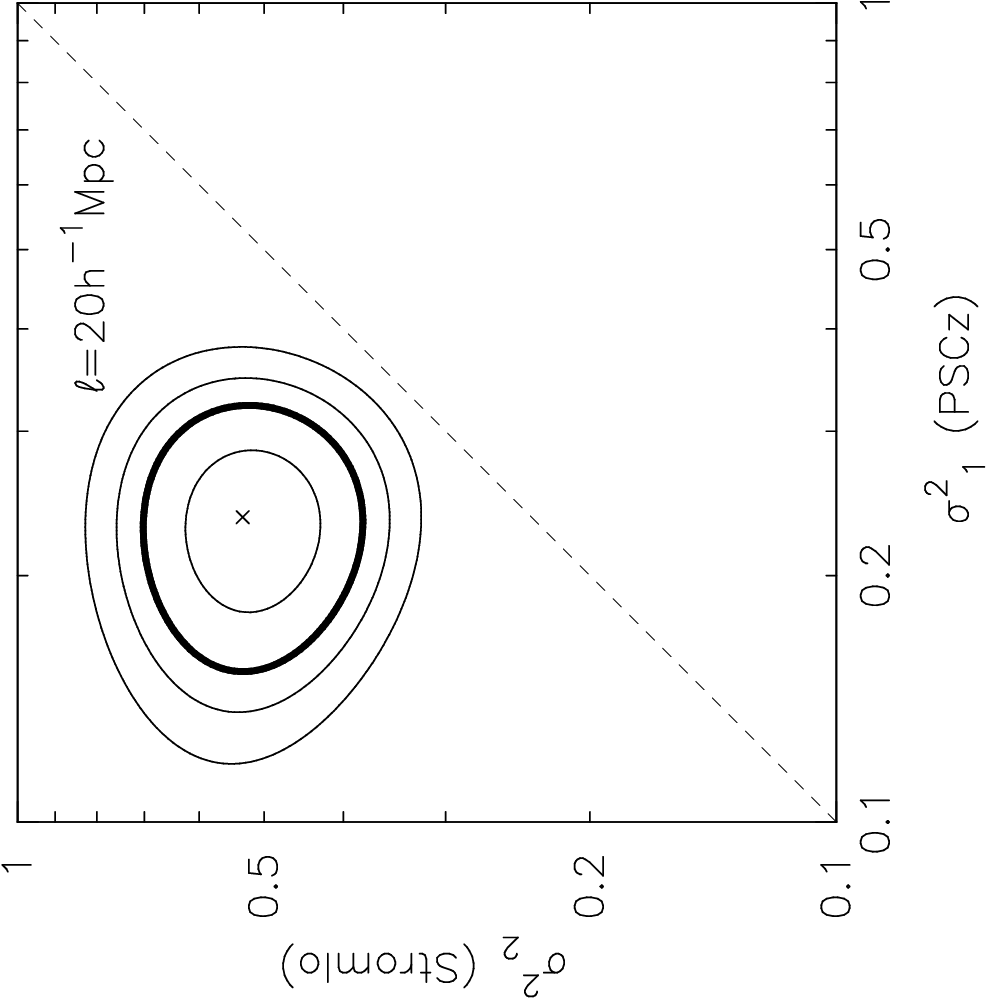}
\includegraphics{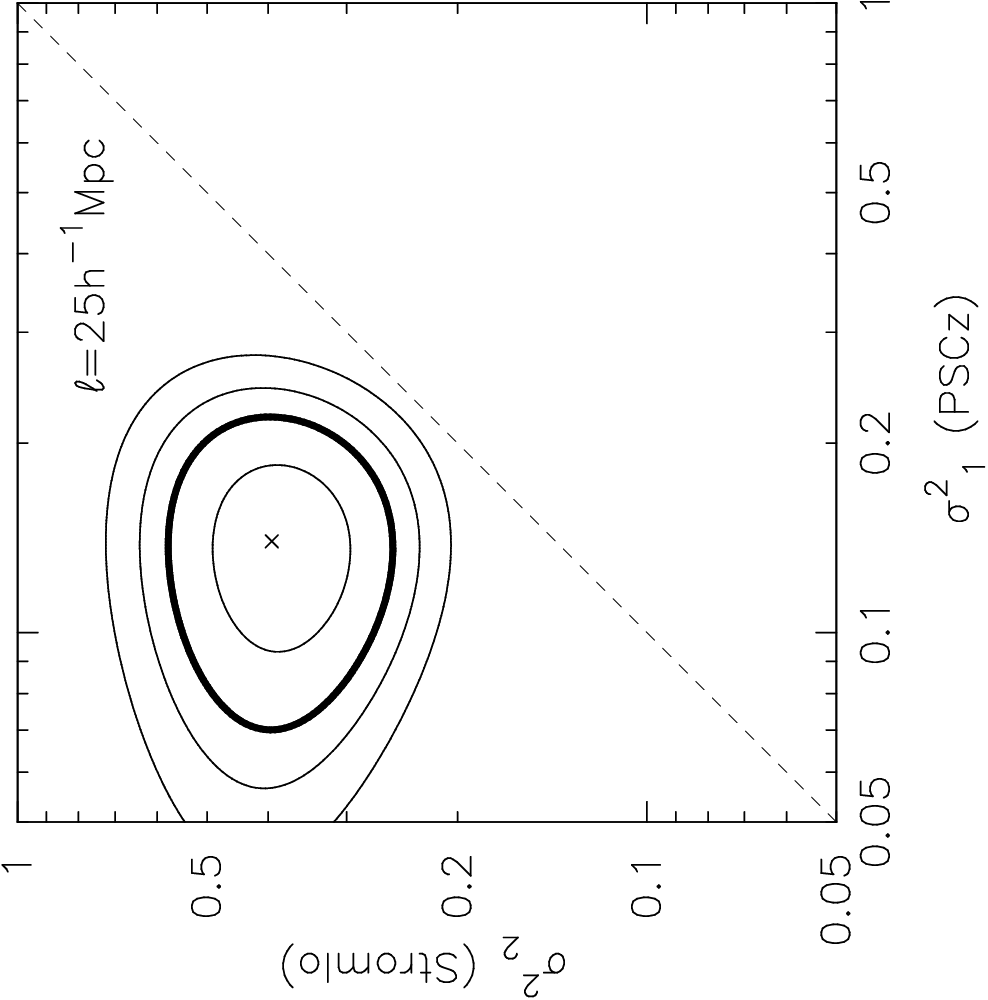}
\includegraphics{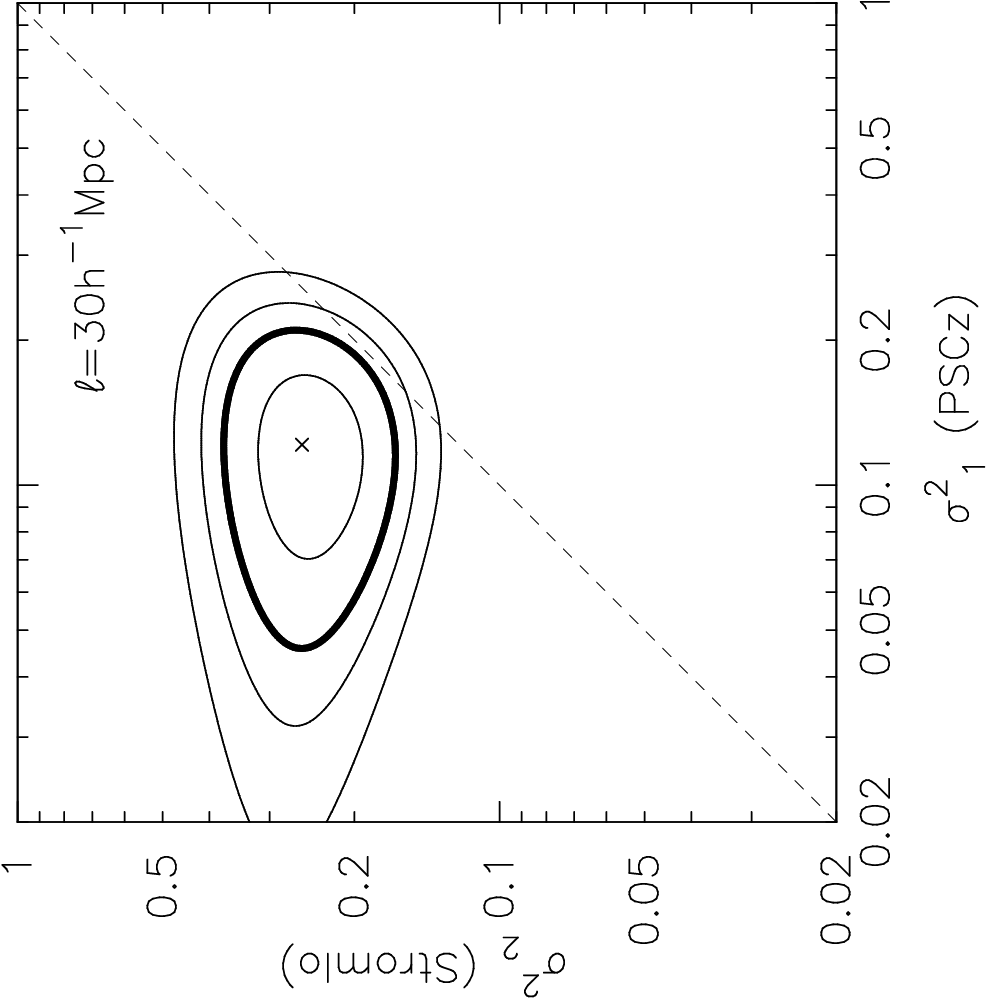}
\includegraphics{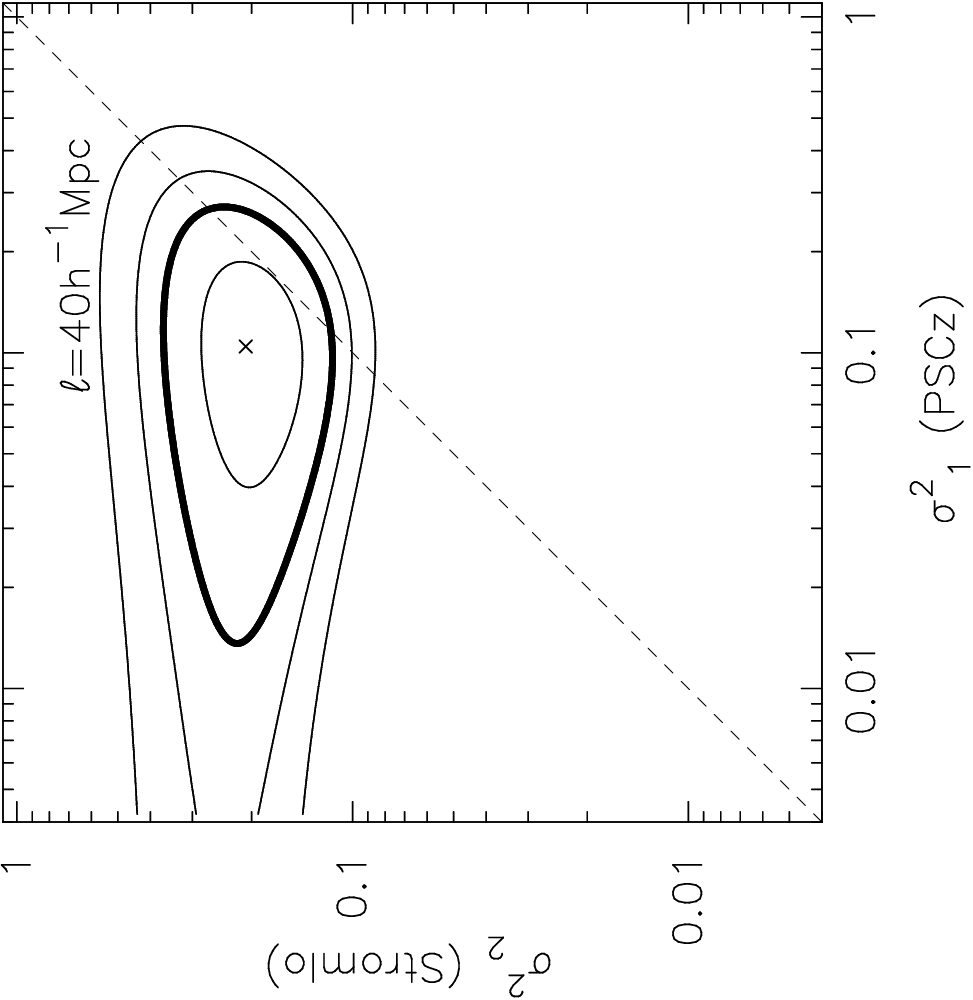}
\vspace*{5mm}

\caption{Contour maps of the likelihood of variances of PSCz and
Stromlo galaxy counts for a range of cell sizes. In each case we have
assumed $\sigma_{12}^2 = \sigma_1\sigma_2$. The contours show where
$2\ln(\like/\likemax)$
is equal to 2.28, 5.99, 9.21 and 13.81. These correspond to 68\%, 95\%
(thick contour), 99\% and 99.9\% confidence intervals respectively,
according to a chi-squared distribution with two degrees of
freedom. The dotted line shows $\sigma_1^2 = \sigma_2^2$  \label{contour}} 
\end{figure*}

Figure \ref{contour} shows contour maps of the joint likelihood 
assuming a linear bias, ${\cal L}(\sigma_1^2,\sigma_2^2, R=1)$, 
for cells of size $10-40 \hmpc$. 
Again we repeat the
analysis using the eight different grid positions,
and compute the geometric mean of the individual ${\cal L}$'s.
This improves our estimate of the peak location but the contours
do not then accurately reflect our improved knowledge of the two
variances. Thus, we compute the value of $\sigma^2$ for the 27 pairs
of mock catalogues, this time using only galaxies that lie outside the
joint mask. We finally rescale the mean $\log {\cal L}$ so the
width of the function matches that in the simulations. 

We approximate the shape of the likelihood function to that produced
by two Gaussian distributions such that the value of
$2\ln(\like / \likemax)$
has a chi-squared distribution with two degrees of freedom. This
approximation was tested by directly determining the value of the
contour that contains 95\% of the likelihood enclosed within a
large area of the $\sigma_1^2$-$\sigma_2^2$ ($R=1$) plane. For each
cell size examined the 95\% contour (shown in bold in figure
\ref{contour}) accurately coincided with the contour
$2\ln(\like / \likemax)=5.99$
which is the 5th percentile point of the $\chi^2$
distribution with two degrees of freedom.

For cells of size $15 - 30 \hmpc$ we rule out the
hypothesis of equal bias as the line 
$\sigma_1^2=\sigma_2^2$
lies well outside the 95\% confidence interval. The variance of
the counts in cells for Stromlo-APM are clearly larger than those for
PSCz. For cells of size 10 and $40 \hmpc$ a unit bias is
marginally consistent with the larger errors, but a value $> 1$ is
still favoured. All the cell sizes are consistent with a 
scale-independent relative bias of $\brel = 
b_{\rm Stromlo} / b_{\rm PSCz} \approx 1.4$. 

\begin{table*}
\centering
\caption{Relative bias as a function of cell size. \label{biastab}}
\vspace{0.1in}
\begin{tabular}{lrc}\hline\hline
	$\ell$
	&\multicolumn{2}{r}{$\brel (95\%)$}\\ \hline
	10	&1.24 &(0.97--1.59)\\
	15 	&1.38 &(1.15--1.68)\\
	20	&1.50 &(1.25--1.86)\\
	25	&1.68 &(1.31--2.34)\\
	30	&1.46 &(1.12--2.30)\\
	40	&1.42 &(0.96--3.30)\\
\end{tabular}
\end{table*}
Table \ref{biastab} shows the relative bias for each cell size. The value
$\brel^2$ is given by the
ratio of the maximum likelihood variances
$\sigma_{\rm{Stromlo}}^2/\sigma_{\rm{PSCz}}^2$. The 95\% confidence
interval was found by determining the two straight lines
$\sigma_{\rm{Stromlo}}^2=b_{\rm{min}}^2\sigma_{\rm{PSCz}}^2$ and
$\sigma_{\rm{Stromlo}}^2=b_{\rm{max}}^2\sigma_{\rm{PSCz}}^2$ on the
contour plots between which 95\% of the total likelihood lies.

\section{The two-point and cross-correlation functions}

The linear bias model predicts that the individual two-point
correlation functions, $\xi_1$ and $\xi_2$, will be related to the
cross-correlation function, $\xi_{12}$, by
\begin{equation}
	\xi_{12}=\left(\xi_1\xi_2\right)^{1/2}.\label{eq:lincorr}
\end{equation}
The three correlation functions were found for the galaxies within the
common survey region using the estimators
\begin{eqnarray}
\nonumber \xi_1(r)=\frac{\langle D_1D_1\rangle \langle R_1R_1\rangle}
    {\langle D_1R_1\rangle^2} -1 \\
 \xi_{12}(r)=\frac{\langle D_1D_2\rangle \langle R_1R_2\rangle}
      {\langle D_1R_2\rangle\langle D_2R_1\rangle}-1
\end{eqnarray}
(Hamilton 1993) where the notation $\langle DD\rangle$, $\langle
RR\rangle$ and $\langle DR\rangle$ refers to the (weighted) number of
data-data, random-random and data-random pairs respectively in a
narrow bin of separation $r$. The subscripts 1 and 2 again correspond
to the two surveys. The two random catalogues were generated within
the joint survey volume, each with the correct corresponding radial
selection function. The selection functions were determined by
integrating the published luminosity functions for the two surveys. For
Stromlo-APM we use a Schechter function with input parameters
$\alpha=-0.97$, $M^\ast=-19.50$ and $\phi^\ast=1.4\times10^{-2}$
$h^3$Mpc$^{-3}$ (Loveday et al.\, 1992). For the \emph{IRAS} luminosity
function we use the parametric form
\begin{equation}
\phi(L)=C\left(\frac{L}{L^\ast}\right)^{(1-\alpha)}
\exp\left[-\frac{1}{2\sigma^2}\log_{10}^2\left(1+\frac{L}{L^\ast}\right)\right]
\end{equation}
with $C=2.6$ $h^3$Mpc$^{-3}$, $\alpha=1.09$, $\sigma=0.724$ and
$L^\ast=10^{8.47}$ $h^{-2}L_{\sun}$ (Saunders et al.\, 1990). Each
random catalogue contains $2\times 10^4$ data points.

To each galaxy or random in a pair we assign the weight
\begin{equation}
	w_i = \frac{1}{1+4\pi n(r_i)J_3(r)}, \ \ \  J_3(r)=
	\int_0^r\xi(x)dx,
\end{equation}
where $r_i$ is the radial distance to the galaxy. This weighting
scheme can be shown to give the minimum uncertainty in $\xi(r)$ on
scales where $\xi(r)\la 1$. We determine the input quantity $J_3$
using determinations of the \emph{IRAS} and Stromlo correlation
functions by previous authors. We use power law fits of the form
$\xi(r)=(r/r_0)^{-\gamma}$. Fisher \etal\ (1994) have measured the
correlation function of the \emph{IRAS} 1.2-Jy survey and find that
their results are well described by a power law with $r_0=4.53$
$h^{-1}$ Mpc and $\gamma=1.28$. The Stromlo-APM correlation function
has been determined by Loveday \etal\ (1995) and they obtain a power
law with $r_0=5.9$ $h^{-1}$ Mpc and $\gamma=1.47$. The 1$\sigma$ error
bars were calculated using the 27 pairs of mock PSCz and Stromlo
catalogues under the same weighting scheme.

\begin{figure}

\hspace{5mm}\raisebox{65mm}{\rotatebox{270}{\scalebox{0.32}[0.29]%
{\includegraphics[110mm,80mm]{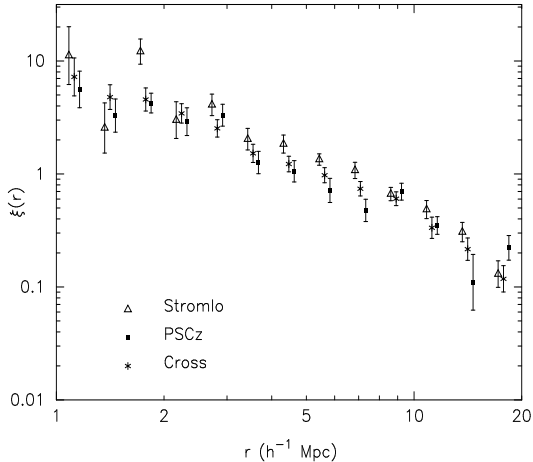}}}}

\caption{The Stromlo, PSCz and cross-correlation functions for the
overlapping volume. 
The $1\sigma$ error bars were generated using the
27 pairs of mock catalogues.\label{corfuns}}
\end{figure}

The three correlation functions are shown in figure \ref{corfuns}.
The Stromlo correlation function is in general greater than that of
PSCz, consistent with the cell count variances. As expected, the cross
correlation function lies between the two individual functions.

\begin{figure}
\hspace{5mm}\raisebox{65mm}{\rotatebox{270}{\scalebox{0.32}[0.29]%
{\includegraphics[110mm,80mm]{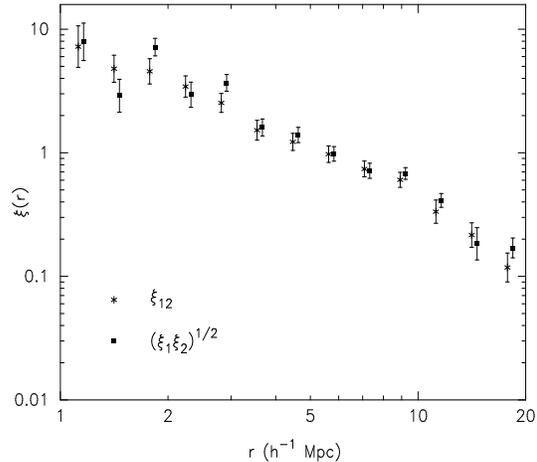}}}}

\caption{The cross correlation function, $\xi_{12}$, for PSCz and
Stromlo-APM and the quantity $(\xi_1\xi_2)^{1/2}$ where $\xi_1$ and
$\xi_2$ are the individual two point correlation functions. The error
bars show one standard deviations calculated from the mock
catalogues.\label{corr12}}
\end{figure}

The cross-correlation function $\xi_{12}$ and the quantity
$(\xi_1\xi_2)^{1/2}$ are shown in Figure \ref{corr12}. 
This shows that these two quantities are consistent within
their error bars, i.e. in good agreement with the linear biasing
hypothesis. If the galaxy distributions were related in some other way
- for example non-linear or stochastic biasing - we would expect
$\xi_{12}$ to lie below $(\xi_1\xi_2)^{1/2}$. We see from
Figure \ref{corr12} that there is no such systematic difference.
Whereas the counts in cells technique was only used
to investigate structure on scales of $\ell\geq 10 \hmpc$, here
we see that the linear bias model is consistent right down to scales of
$\sim 1.5 \hmpc$. On scales smaller than this our determination
of the three correlation functions becomes highly uncertain as there
are a low number of galaxy pairs at these separations.

\begin{figure}

\hspace{5mm}\raisebox{65mm}{\rotatebox{270}{\scalebox{0.32}[0.29]%
{\includegraphics[110mm,80mm]{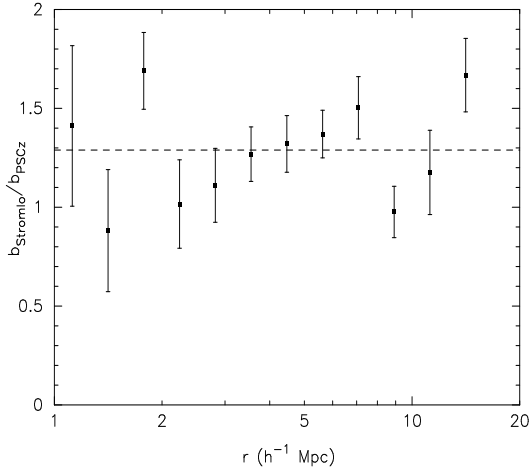}}}}

\caption{The relative bias between the Stromlo and PSCz correlation
functions  over the range $1 - 20 \hmpc$. The dotted line
is the mean value
$b_{\rm{Stromlo}}/b_{\rm{PSCz}}=1.29$\label{biascorr}}
\end{figure}

The relative bias $\brel \equiv b_{\rm{Stromlo}}/b_{\rm{PSCz}}$ is
estimated from the correlation
functions at distances in the range $1-20 \hmpc$ using
\begin{equation}
    \brel  = \left(\frac{\xi_{\rm{Stromlo}}}{\xi_{\rm{PSCz}}}\right)^{1/2}
\end{equation}
and is plotted in Figure \ref{biascorr}.
Beyond separations of $r \approx 30 \hmpc$ the relative bias becomes
highly uncertain as both correlation functions are consistent with
zero on these scales. Where the $1\sigma$ error bars are not too
large, we see that the relative bias remains roughly constant over the
separation range $r < 20 \hmpc$ and has a mean value of $\brel = 1.29
\pm 0.07$, where the error is estimated from the mock catalogues.  A
calculation of $\chi^2=\sum_{i=1}^N {(b_i-\bar{b})^2/\sigma_i^2}$
gives a reduced chi-squared of 1.76. Although a little high, this
value does not rule out the hypothesis that the relative bias is
unchanging with scale within a 95\% confidence limit. (Note that
caution should be taken in the interpretation of this $\chi^2$ value
since the measurements of the values of $b$ on different scales are
not independent.) More importantly, we note that there is no obvious
trend in the relative bias with scale.

We estimate the average correlation coefficient $R = \xi_{12} /
\sqrt{\xi_1\xi_2}$ by taking an unweighted mean over data points in
the range $2 \hmpc < r < 20 \hmpc$; the result is $R = 0.93 \pm 0.06$,
where we again derive the $1\sigma$ error from the simulations. This
can be used to place limits on ``stochastic bias'' as discussed later.

\section{The Tegmark `null-buster' test}

\begin{figure*}
\vspace*{17.0cm}
\includegraphics{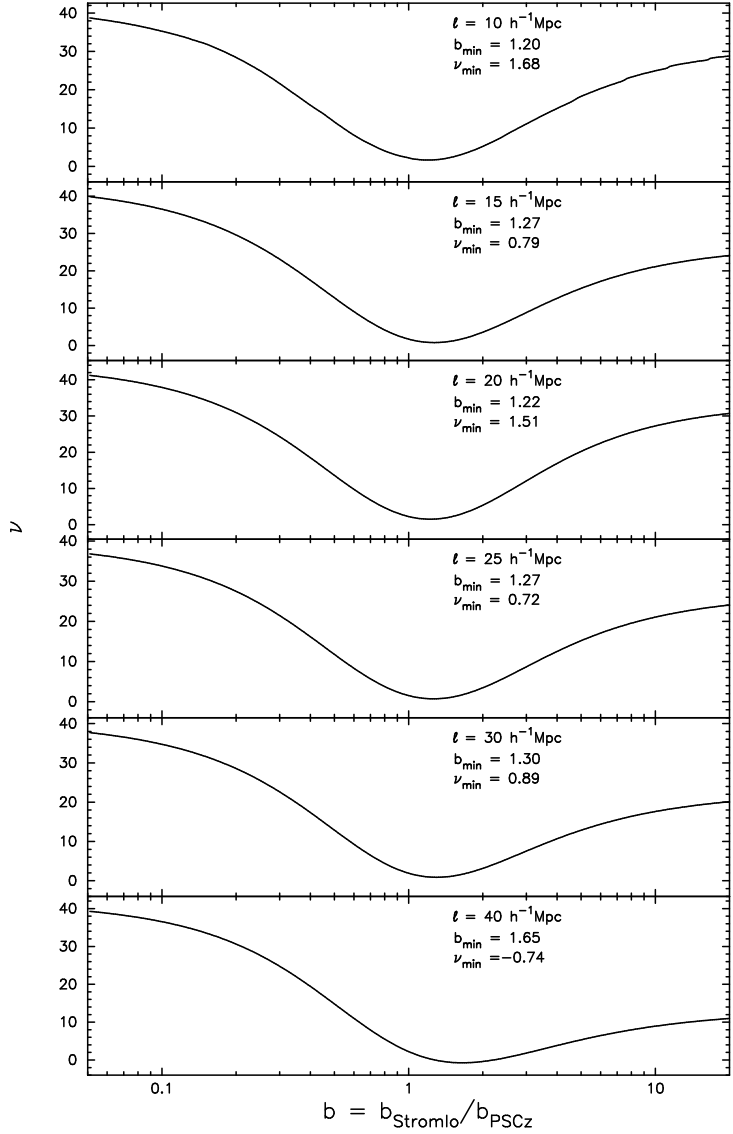}

\caption{The function $\nu(\brel)$, 
i.e. number of ``sigmas'' at which the linear bias
$g_{\rm Stromlo} = \brel g_{\rm PSCz}$ can be ruled out
as a function of $\brel$. The position of the minimum is given for each
cell size. \label{nullbust}}
\end{figure*}

Tegmark and Bromley (1998) use a generalised $\chi^2$-statistic to
directly compare cell counts of different galaxy populations in
the Las Campanas Redshift Survey. We have applied their method to compare
the clustering in PSCz and Stromlo-APM. 
We bin the galaxies from the
two surveys using the cells described earlier and, for each of $\Nc$
cells, we calculate the overdensities $g_i^{\rm{(PSCz)}} =
N_i^{\rm{(PSCz)}}/\langle N\rangle^{\rm{(PSCz)}}-1$ and
$g_i^{\rm{(Stromlo)}} = N_i^{\rm{(Stromlo)}}/\langle
N\rangle_i^{\rm{(Stromlo)}}-1$ ($i=1,\ldots,\Nc$). The expected
counts $\langle N\rangle_i^{\rm{(PSCz)}}$ and $\langle
N\rangle_i^{\rm{(Stromlo)}}$ were calculated using the respective
selection functions and the joint survey mask. If the galaxy densities
are related by linear bias, then there will exist a value for
$\brel$ such that the (column) vector
\begin{equation}
   \Delta g = g_{\rm{Stromlo}} - \brel g_{\rm{PSCz}}
\end{equation}
is consistent with shot noise. 
The covariance matrix of $\Delta g$ is diagonal and
(for density fluctuations $\simlt 1$) is given by 
\begin{equation}
	\mathbf{N}\equiv\frac{1}{\langle N\rangle_{\rm{Stromlo}}} +
	\frac{\brel^2}{\langle N\rangle}_{\rm{PSCz}}.
\end{equation}
Our null hypothesis is that, for a given value of $b$, $\langle\Delta
g\rangle=0$ and $\langle\Delta g\Delta g^t\rangle=\mathbf{N}$. We might
choose to test this hypothesis by calculating the statistic
$\chi^2=\Delta g^t\mathbf{N}^{-1}\Delta g$. The number of ``sigmas''
at which the null hypothesis is ruled out is then
$\nu=(\chi^2-\Nc)/\sqrt{2\Nc}$. 
If we have an alternative hypothesis
that there is an extra signal with covariance matrix $\mathbf{S}$, such that
$\langle\Delta g\Delta g^t\rangle= \mathbf{N}+\mathbf{S}$, 
then Tegmark (1998) shows that the statistic $\Delta
g^t\mathbf{N}^{-1}\mathbf{S}\mathbf{N}^{-1}\Delta g$ will provide a
more sensitive test (by using our prior knowledge 
of the signal covariance matrix). 
 The significance at which the null hypothesis can
be ruled out is then given by
\begin{equation}
	\nu\equiv\frac{\Delta
	g^t\mathbf{N}^{-1}\mathbf{S}\mathbf{N}^{-1}\Delta g - \mbox{tr}
	\mathbf{N}^{-1}\mathbf{S}}{[2\mbox{tr}
	\{\mathbf{N}^{-1}\mathbf{S}\mathbf{N}^{-1} \mathbf{S}\}]^{1/2}}.
\end{equation}
If the biasing is non-linear then the deviations from linearity would
be correlated with large scale structure. We thus choose the matrix
$\mathbf{S}$ to be the covariance between cell overdensities
calculated using the PSCz correlation function, i.e.  the
volume-average of $\xi(r_{ij})$ over cells $i$ and $j$.  (Note that
$\nu$ depends only on the shape of $\mathbf{S}$, not its amplitude).
The resulting values of $\nu$ as a function of $b$ for a range of cell
sizes are shown in Figure~\ref{nullbust}.

As expected we see that extreme values of the relative bias parameter
are ruled out with high significance (i.e., 
the clustering in each individual survey is 
inconsistent with pure shot noise). 
The position of the minimum indicates the most 
likely value of $\brel$: Table \ref{tegtab} shows this
value for each cell size. The $1\sigma$ errors
are estimated using the 27 simulated catalogues from
\S~\ref{sec-sims}. 
For all cell sizes we see that the minimum value 
of $\nu$ is typically not much larger than $1$, again 
suggesting that the two surveys are consistent with a linear relative
bias. At all scales the relative bias is consistent with 
$\brel \approx 1.3$, 
again above unity but slightly lower than the values estimated in 
\S3. 

\begin{table*}
\centering
\caption{Relative bias as a function 
of cell size.\label{tegtab}}
\vspace{0.1in}

\begin{tabular}{lc}
\hline\hline
 $\ell$  & $\brel$ \\ 
 \hline
	15 	& $1.27 \pm 0.11$ \\
	20	& $1.22 \pm 0.09$ \\
	25	& $1.27 \pm 0.12$ \\
	30	& $1.30 \pm 0.16$ \\
	40	& $1.65 \pm 0.13$ \\
\hline
\end{tabular}

\end{table*}

\section{Conclusions}

We have compared the clustering of two redshift surveys
(the new \emph{IRAS}-selected PSCz survey, and the optically-selected Stromlo-APM 
survey) within their common region of space. 
Three  complementary methods have been used: the counts-in-cells
method, the two-point correlation function, and the  Tegmark `null-buster'
test. 
In all three cases the results  are consistent with  a  linear
biasing model, i.e. $\delta_{\rm Stromlo} = \brel \delta_{\rm PSCz}$
with  $\brel = b_{\rm Stromlo}/b_{\rm PSCz} \approx 1.3 \pm 0.1$. 
There is little evidence for variation of the relative
bias over the range of scales from $\sim 5 - 30 \hmpc$.  
We find a lower limit on the correlation  coefficient $R \simgt 0.85$
on scales $\sim 10 - 20 \hmpc$. 

Our value for the relative bias is in quite good agreement with
earlier estimates; Baker \etal\ (1998) found $\brel = 1.4$; Willmer
\etal\ found $\brel = 1.20\pm0.07$ and Saunders \etal\ (1992) obtained a
value of $\brel = 1.38\pm0.12$ in real space.  Our high value for $R$
seems somewhat unexpected in view of the results of Tegmark \& Bromley
(1998), who found values of $R$ as low as $0.5$ comparing various
pairs of spectral classes in the Las Campanas redshift survey.
However, their results are not directly comparable with ours for
several reasons: firstly, they used considerably smaller cells of size
$\sim 6 \hmpc$, and secondly our galaxy classes (PSCz and Stromlo) do
not correspond closely to the classes used by TB; Stromlo contains a
mix of early and late type galaxies, while PSCz is weighted towards
late-type spirals with a preference for high surface brightnesses
(hence higher dust temperatures) and merging systems.

We may translate a lower limit on  $R$ into an upper limit 
 on `stochastic' bias {\sl if} it is independent between the two 
 galaxy classes, as follows: using the notation of Dekel \& Lahav
(1998), we may define the stochasticity  $\epsilon_i (i=1,2)$ by 
 $g_i = \langle g_i \vert \delta \rangle + \epsilon_i$
where the angle brackets denote averages over $\delta$. 
Then 
\begin{eqnarray}
{\rm Cov}(g_1,g_2) & = & \langle \langle g_1\vert\delta \rangle
  \langle g_2\vert\delta \rangle \rangle + 
 \langle \epsilon_1 \epsilon_2 \rangle \nonumber 
 \\ 
  & \equiv & \sigma^2 (\btil_{12}^2 + \sigma_{b12}^2)  \ \ {\rm thus}   
 \\
 R & \equiv & {\rm Cov}(g_1,g_2) / \sigma_{g1} \sigma_{g2} \nonumber 
 \\
  & = & {\btil_{12}^2 + \sigma_{b12}^2 \over 
     \left[ (\btil_1^2 + \sigma_{b1}^2) (\btil_2^2 + \sigma_{b2}^2) 
    \right]^{1/2}  }
\end{eqnarray}  
where the first line follows since the cross terms 
$\langle \langle g_1\vert\delta \rangle \epsilon_2\rangle$ vanish , 
and the second line follows from 
the definitions of $\btil_{12}$ and
$\sigma_{b12}$, analogous to the definitions
$\btil_1^2 \equiv 
\langle \langle g_1 \vert\delta \rangle^2 \rangle / \sigma^2$ 
and $\sigma_{b1}^2 \equiv \langle \epsilon_1^2 \rangle / \sigma^2$
in DL, and $\sigma^2$ is the mass variance on a given smoothing scale. 
From numerical studies, DL find that $\btil / \hat{b} \sim 1 - 1.15$, 
 so it is probably reasonable to assume 
$\btil_{12} \approx (\btil_1\btil_2)^{1/2}$. 
Then if we assume 
$\sigma_{b12} = 0$ (i.e. assuming that the `stochastic' bias
is independent between surveys),  and adopting 
$\btil_{\rm PSCZ} = 1, \btil_{\rm Stromlo} = 1.3$, we find that  
$R \ge 0.8$ would imply $\sigma_{b1}^2 < 0.56$ 
and $\sigma_{b2}^2 < 0.95$,
 Likewise $R \ge 0.9$ would imply
 $\sigma_{b1}^2 < 0.23$ and $\sigma_{b2}^2 < 0.39$.  
DL show that $b_{\rm var}^2 = \btil^2 + \sigma_b^2$, so
the ``fractional stochasticity'' is effectively $\sigma_b^2 / \btil^2$,
i.e. the ratio of stochastic variance to deterministic variance
in galaxy density.  
If the fractional stochasticity is equal for the two surveys, such that
$\sigma_{b1}^2 / \btil_1^2 = \sigma_{b2}^2 / \btil_2^2 = y$, 
then the above assumptions lead to $1 + y = R^{-1}$. 
Clearly, these numbers are illustrative rather than 
definitive: 
the assumption $\sigma_{b12}^2 = 0$ may not be satisfied
in practice since there could be a `second parameter' in addition
to the mass overdensity  
e.g. local shear, gas temperature etc. which affects the formation 
of both \emph{IRAS} and optical galaxies. But this argument does suggest 
that `stochastic' bias which preferentially affects either
the earliest or latest type galaxies is unlikely to be severe. 

Our results are encouraging for velocity field studies, in that 
that they suggest that large-scale density fields
(usually estimated from \emph{IRAS} galaxies) should also
be a good match to those which would be estimated from 
all-sky optical surveys (currently rather limited in extent); 
this supports the conclusions of Baker \etal\ (1998). 

In a future paper we plan to apply the tests considered here
to the PSCz and the Optical Redshift Survey (Santiago \etal\ 1995), 
which gives a higher sampling density than Stromlo-APM and thus
may provide stronger constraints. 

\section*{Acknowledgements}

We are very grateful to many observers, especially Marc Davis, Tony
Fairall,
Karl Fisher \& John Huchra for supplying redshifts
in advance of publication.
We thank the staff at the INT, AAT and CTIO telescopes for support,
especially Hernan Tirado and Patricio Ugarte at CTIO.
MDS is supported by a PPARC studentship, 
W.~Sutherland is supported by a PPARC Advanced Fellowship, 
and W.~Saunders is supported by a Royal Society Fellowship.
GE thanks PPARC for the award of a Senior Fellowship.

\end{document}